\documentclass[10pt,journal,compsoc]{IEEEtran}
\ifCLASSOPTIONcompsoc
  \usepackage[nocompress]{cite}
\else
  \usepackage{cite}
\fi

\usepackage{amsmath,amssymb,amsfonts}
\usepackage{graphicx}
\usepackage{textcomp}
\usepackage{xcolor}
\usepackage{comment}
\usepackage{xspace}
\usepackage{pifont}
\usepackage{multirow}
\usepackage{booktabs}
\usepackage{colortbl}
\usepackage{tcolorbox}
\usepackage{ulem}
\usepackage{fvextra}
\usepackage{enumitem}
\usepackage{listings}
\usepackage{cases}
\usepackage{url}
\usepackage[ruled,linesnumbered,lined,noend]{algorithm2e}
\usepackage{array}

\ifCLASSOPTIONcompsoc
  \usepackage[caption=false,font=footnotesize,labelfont=sf,textfont=sf]{subfig}
\else
  \usepackage[caption=false,font=footnotesize]{subfig}
\fi

\usepackage[hidelinks]{hyperref}

\definecolor{light-gray}{gray}{0.95}

\newcommand{\tool}{\textsf{RTL-Obliger}}
\newcommand{\bench}{\textsc{SecRTL-Gen}}

\definecolor{dkgreen}{rgb}{0,0.6,0}
\definecolor{gray}{rgb}{0.5,0.5,0.5}
\definecolor{mauve}{rgb}{0.58,0,0.82}

\begin{document}

\title{Unsaid, Unsafe? Implicit Security Obligations in LLM-Based RTL Code Generation}

\author{Guang~Yang,
Xing~Hu\IEEEauthorrefmark{1},
Xiang~Chen,
and~Xin~Xia%
\IEEEcompsocitemizethanks{%
\IEEEcompsocthanksitem G.~Yang is with the State Key Laboratory of Blockchain and Data Security, Zhejiang University, Hangzhou, China, and also with the Hangzhou High-Tech Zone (Binjiang) Institute of Blockchain and Data Security, Hangzhou, China.
\IEEEcompsocthanksitem X.~Hu and X.~Xia are with the State Key Laboratory of Blockchain and Data Security, Zhejiang University, Hangzhou, China.
\IEEEcompsocthanksitem X.~Chen is with the School of Artificial Intelligence and Computer Science, Nantong University, Nantong, China.
\IEEEcompsocthanksitem Corresponding author: Xing Hu (e-mail: xinghu@zju.edu.cn).
Other e-mails: novelyg@outlook.com, xin.xia@acm.org, xchencs@ntu.edu.cn.}%
\thanks{Manuscript received XXX XX, 2026; revised XXX XX, 2026.}}

\markboth{IEEE Transactions on Dependable and Secure Computing}%
{Yang \MakeLowercase{\textit{et al.}}: Unsaid, Unsafe? Implicit Security Obligations in LLM-Based RTL Generation}

\IEEEtitleabstractindextext{%
\begin{abstract}
Large Language Models (LLMs) generate register-transfer-level (RTL) code with rapidly improving functional correctness. Security of LLM-generated code, however, has been studied mainly for software, where flaws can still be patched after deployment. Insecure RTL offers no such remedy once taped out into silicon.
We construct {\bench}, a multi-language \emph{resource-access} security benchmark grounded in real SoC IP: 392 tasks over five CWE families and four HDLs (Verilog, SystemVerilog, VHDL, and Python), each with black-box functional and security testbenches. Functional specifications intentionally omit security obligations, matching how obligations are often kept out of functional docs in practice.
An empirical study of five frontier LLMs shows a sharp gap: under vanilla prompts they pass functional tests in about 73--79\% of cases but security tests in only 14--35\%, and stronger functional models are not safer. Adding CWE knowledge raises security, while unaided self-thinking helps less and both security-oriented prompts cut functional pass rates, showing that the bottleneck is missing weakness awareness in the specification, not an inability to write defensive RTL.
We present \tool{}, a neuro-symbolic framework that infers these implicit obligations. An LLM extracts a functional-semantic graph from the specification; a symbolic engine then matches it against a CWE pattern ontology to surface mitigation-evidence gaps and signal-level obligations; the LLM finally revises RTL under those obligations in a functionality-preserving two-stage generation.
Across five models and four languages, \tool{} raises mean all-pass from 49.6--51.4\% (SecV/RESCUE) to 61.6\%, with higher security and functional rates than these secure-generation baselines.
\end{abstract}

}

\maketitle
\IEEEdisplaynontitleabstractindextext
\IEEEpeerreviewmaketitle

\IEEEraisesectionheading{\section{Introduction}\label{sec:intro}}

\IEEEPARstart{L}{arge} language models (LLMs) are increasingly used to generate register-transfer-level (RTL) code~\cite{yang2025large,liu2023verilogeval,lu2024rtllm}.
Functional pass rates keep rising, but security has received far less attention.
This gap matters more for hardware than for software.
Software vulnerabilities can often be patched after deployment.
Insecure RTL is baked into silicon after tape-out and cannot be fixed in the field.

The security of LLM-generated software is already a known concern~\cite{pearce2022asleep,tony2023llmseceval}.
Hardware raises a sharper question: given only a functional specification, does the model also enforce the defenses a safe design needs?
In practice, that specification states what the module should do, while many required defenses live in separate security configurations, threat models, or review checklists.
We call these \textbf{implicit security obligations}: constraints still required of a safe implementation, yet not entailed by the written functional specification (\S\ref{sec:bg-implicit}).
Human designers can recover them from those external sources; but the LLM sees only the functional text, so the obligations never enter its input.
OpenTitan illustrates the pattern: countermeasures appear in machine-readable configuration files, not in the functional IP description~\cite{opentitan2025comportability,meza2023opentitan}.
Standard benchmark-style prompts go further: they typically contain no security-obligation content at all.

Existing work does not close this gap.
RTL generation benchmarks mainly score functional simulation~\cite{liu2023verilogeval,lu2024rtllm,pinckney2025comprehensive}.
Recent security benchmarks and audits show that LLM-written RTL can pass functional checks and still fail security tests~\cite{chen2026hardsecbench,ibnat2025trusting}, but they do not focus on real SoC IP under multi-HDL, obligation-omitted prompts.
Secure-generation methods such as SecV~\cite{fan2025secv} and RESCUE~\cite{shi2026rescue} inject CWE knowledge through graph exploration or retrieval.
They still rely on the LLM to turn coarse guidance into design-specific constraints, which is exactly where our controlled study finds the bottleneck (\S\ref{sec:empirical}).

This paper asks two linked questions.
First, how large is the functional--security gap when LLMs generate RTL from security-incomplete specifications?
Second, can we infer the missing obligations and use them to generate secure RTL without breaking functionality?
We scope both questions to \emph{resource-access} weaknesses that are port-observable under black-box simulation, matching the families covered by {\bench}.

We build {\bench} (\S\ref{sec:bench}) to measure the gap.
Empirically, many implicit obligations appear as mitigation-evidence gaps: the specification shows weakness preconditions (for example, a reusable secret register is written) but not the required mitigation.
If the specification is lifted into a checkable evidence model, and complete mitigation patterns are stored in a reusable ontology, obligation inference becomes a deterministic signal-level check rather than open-ended guessing.
We therefore present \tool{} (\S\ref{sec:method}): an LLM extracts a functional-semantic graph; a symbolic engine matches it against a CWE pattern ontology to emit signal-level obligations; and the LLM revises RTL in two stages, first a functional draft and then local obligation-guided edits.
On {\bench}, five frontier LLMs under vanilla prompts pass functional tests in 73--79\% of cases but security tests in only 14--35\%; \tool{} raises mean all-pass from 49.6--51.4\% (SecV/RESCUE) to 61.6\%.

In summary, we make four contributions:
\begin{itemize}[leftmargin=*,nosep]
  \item \textbf{{\bench}:} a 392-task, four-HDL security benchmark from real SoC IP, with black-box functional and security tests under obligation-omitted specifications (\S\ref{sec:bench}).
  \item \textbf{Empirical study:} five LLMs reach 73--79\% functional pass but only 14--35\% security pass; CWE knowledge helps more than self-thinking, and security prompting lowers functional pass (\S\ref{sec:empirical}).
  \item \textbf{\tool{}:} a neuro-symbolic method that infers signal-level obligations from a functional-semantic graph and a CWE ontology, then applies them in two-stage RTL generation (\S\ref{sec:method}).
  \item \textbf{Evaluation:} \tool{} raises mean all-pass from 49.6--51.4\% (SecV/RESCUE) to 61.6\%; symbolic inference is the main gain, and residual failures center on obligation-guided generation and matching recall (\S\ref{sec:rq1-baselines}--\S\ref{sec:rq3-errors}).
\end{itemize}

To facilitate the replication of {\tool}, we make our datasets and source code publicly available on GitHub.\footnote{\url{https://github.com/NTDXYG/SecRTL-Gen}}

\medskip

\noindent\textbf{Paper structure.} Section~\ref{sec:background} reviews LLM-based RTL generation, hardware security weaknesses, and implicit security obligations. Section~\ref{sec:empirical} presents {\bench} and the empirical study of the functional--security gap. Section~\ref{sec:method} introduces the \tool{} framework. Section~\ref{sec:results} reports evaluation results. Section~\ref{sec:discussion} discusses implications and threats to validity, Section~\ref{sec:related} reviews related work, and Section~\ref{sec:conclusion} concludes.
\section{Background}
\label{sec:background}

\subsection{LLM-Based RTL Code Generation}
\label{sec:bg-codegen}

We formalize LLM-based RTL code generation as a mapping from a natural-language functional specification to RTL code:
\begin{equation}\label{eq:codegen}
  G_\theta: \mathcal{S} \rightarrow \Delta(\mathcal{C}), \quad c \sim G_\theta(s)
\end{equation}
where $\mathcal{S}$ is the space of functional specifications, $\mathcal{C}$ is the space of RTL programs, and $c \sim G_\theta(s)$ is one generation sample.
Functional correctness is checked by a simulation testbench $T_\mathit{func}$; we write $c \models T_\mathit{func}$ when all functional assertions pass.

\subsection{Hardware Security Weaknesses}
\label{sec:bg-cwe}

The Common Weakness Enumeration (CWE)~\cite{mitre2024cwe} catalogs recurring hardware design flaws.
We scope this paper with MITRE's Most Important Hardware Weaknesses (MIHW)~\cite{mitre2021mihw,mitre2025mihw}, and for each selected family $w$ write $\mathcal{O}_w$ for the abstract class of security obligations that an implementation must satisfy to be free of $w$:
\begin{equation}\label{eq:obligation}
  c \text{ is free of } w \;\;{\Longleftrightarrow}\;\; \forall\, o \in \mathcal{O}_w,\;\; c \vDash o
\end{equation}
We derive $\mathcal{O}_w$ from MITRE mitigation guidance.
Equation~\ref{eq:obligation} is definitional: CWE does not provide an exhaustive mitigation set.

\textbf{Example: $\mathcal{O}_{226}$.}
CWE-226 covers sensitive data left in a reusable resource.
For a module that stores secrets, $\mathcal{O}_{226}$ includes: ($o_1$) overwrite every secret-holding element before completion; ($o_2$) do so on error and abort as well as success; ($o_3$) do not expose residual values on ports between completion and the next write.
None of these is needed for functional ciphertext output, so a functional specification can omit them without that omission being visible in $T_\mathit{func}$.

In evaluation, we approximate Equation~\ref{eq:obligation} with a security testbench $T_\mathit{sec}$ over obligations from $\mathcal{O}_w$, and write $c \models T_\mathit{sec}$ when all security assertions pass.
Passing $T_\mathit{sec}$ is evidence for the exercised obligations, not a proof that every $o \in \mathcal{O}_w$ holds.

\subsection{Implicit Security Obligations}
\label{sec:bg-implicit}

In practice, many obligations in $\mathcal{O}_w$ are absent from the functional specification given to a generator~\cite{mohr2023leave,dessouky2019hardfails,riaz2017implicit}.
Let $\mathit{content}(s)$ be the constraints stated in $s$.
We say $s$ is \textit{security-incomplete} w.r.t.\ $w$ when
\begin{equation}\label{eq:implicit}
  \mathit{Incomplete}(s, w) \;\iff\; \exists\, o \in \mathcal{O}_w \;\text{s.t.}\; \mathit{content}(s) \nvDash o
\end{equation}
Entailment is the right test: stating that a buffer must read as zero after completion entails $o_1$ of $\mathcal{O}_{226}$ even without the word ``security''.
The omitted obligations remain required of any implementation that is free of $w$.
This mismatch between what $s$ states and what freedom from $w$ requires is the problem studied in \S\ref{sec:empirical}.

\section{Empirical Study}
\label{sec:empirical}

Existing RTL generation benchmarks~\cite{liu2023verilogeval,lu2024rtllm,pinckney2025comprehensive} check functional correctness, but not security.
We therefore construct {\bench}, a multi-language security benchmark built from real SoC IP, and run a controlled study of this security gap.

\subsection{{\bench}: Benchmark Construction}
\label{sec:bench}

\subsubsection{Scope and CWE Families}
\label{sec:bench-cwe}

Hardware weaknesses span a broad spectrum, from side-channel leakage to power-management flaws.
From the MIHW entries and the broader CWE hardware view~\cite{mitre2024cwe}, we select five CWE families that cover the main resource-access attack surfaces in SoC designs.
Table~\ref{tab:cwe-families} lists the selected families and their case counts.

We scope {\bench} to resource-access security for three reasons.
First, these families are prioritized by MIHW: three were on the 2021 list and all five are on the 2025 list~\cite{mitre2021mihw,mitre2025mihw}.
Second, they are port-observable and thus checkable by black-box simulation.
Third, access-control obligations are often missing from functional specs: security properties are frequently incomplete or kept separate from functional intent~\cite{dessouky2019hardfails,gao2019verisketch,riaz2017implicit}.

\begin{table}[t]
  \caption{CWE families covered by {\bench} (names from MITRE CWE~\cite{mitre2024cwe}). Each family maps to an abstract obligation class $\mathcal{O}_w$.}
  \label{tab:cwe-families}
  \centering
  \small
  \begin{tabular}{@{}l>{\raggedright\arraybackslash}p{0.6\columnwidth}r@{}}
    \toprule
    CWE & Description & \#Cases \\
    \midrule
    CWE-226  & Sensitive Information in Resource Not Removed Before Reuse & 26 \\
    CWE-1189 & Improper Isolation of Shared Resources on System-on-a-Chip (SoC) & 23 \\
    CWE-1256 & Improper Restriction of Software Interfaces to Hardware Features & 26 \\
    CWE-1260 & Improper Handling of Overlap Between Protected Memory Ranges & 7 \\
    CWE-1262 & Improper Access Control for Register Interface & 16 \\
    \midrule
    \multicolumn{2}{@{}l}{Total unique designs ($\times\,4$ HDLs $= 392$ instances)} & 98 \\
    \bottomrule
  \end{tabular}
\end{table}

\subsubsection{Case Construction}
\label{sec:bench-tasks}

Each case is extracted from a real SoC IP block rather than a synthetic example.
We source designs from four open-source projects: OpenTitan~\cite{meza2023opentitan}, Hack@DAC~2021~\cite{chen2022hackdac}, CVA6/Ariane~\cite{zaruba2019cva6}, and PULP Platform~\cite{pullini2019pulp}.
We select modules that meet three criteria: (i)~the module can be compiled and simulated in isolation, without the full SoC bus environment; (ii)~its security property is observable through port-level I/O; and (iii)~it aligns with at least one of the five CWE families.

Each case consists of four artifacts:
\begin{itemize}[leftmargin=*,nosep]
  \item A \textbf{functional specification} describing ports, reset behavior, and functional truth tables or FSM rules, following the prompt style of prior RTL benchmarks~\cite{liu2023verilogeval,lu2024rtllm}.
  \item A \textbf{golden implementation} that satisfies both functional and security requirements, preserved from the original SoC source.
  \item A \textbf{functional testbench} that verifies input/output correctness through simulation assertions.
  \item A \textbf{security testbench} that exercises the security obligations $\mathcal{O}_w$ relevant to the case's CWE family.
\end{itemize}

The functional specification uses a \textit{practice-motivated omission}: it states \textit{what} the module should do, and intentionally withholds how it should be secured.
This instantiates the security-incompleteness notion in \S\ref{sec:bg-implicit}; \S\ref{sec:bench-omission} reports how closely it matches the original documents.
Each case is written in four hardware description languages (Verilog, SystemVerilog, VHDL, and Python via Amaranth), yielding $98 \times 4 = 392$ language-specific instances.
The multi-HDL setup is a first-class part of the benchmark: the same security property must hold across languages.

\subsubsection{Where Security Obligations Live in Practice}
\label{sec:bench-omission}

Our practice-motivated omission matches how real SoC projects document security.
We study OpenTitan as the strongest case among our sources.
Its standard requires every IP to list security countermeasures in a machine-readable configuration file, checked at signoff~\cite{opentitan2025comportability}.
Of the 31 IPs with such a file, 28 list countermeasures (283 in total).
All 283 entries appear in the configuration file; none appear only in the functional description.

This gap matters for RTL generation.
A generation task usually gives the model a functional specification, not the machine-readable configuration file.
So most security obligations never reach the model.
Hack@DAC~2021, CVA6, and PULP do not keep such a configuration record at all.

\subsubsection{Testbench Design}
\label{sec:bench-tb}

\textbf{Functional testbenches:}
we apply stimulus vectors to the design under test (DUT) and check output responses against expected values.
A generated module $c$ passes if $c \models T_\mathit{func}$, i.e., all assertions hold under all stimulus sequences.

\textbf{Security testbenches:}
we exercise scenarios targeting the obligations $\mathcal{O}_w$ of the case's CWE family.
For CWE-226, this means reading a buffer after a transaction completes to detect residual sensitive data; for CWE-1262, issuing an access from an unauthorized privilege level; for CWE-1260, configuring overlapping memory regions to probe priority-enforcement logic.
A module passes if $c \models T_\mathit{sec}$, i.e., the design correctly prevents every exercised attack vector.
As noted in \S\ref{sec:bg-cwe}, $T_\mathit{sec}$ is an operational approximation of checking $\mathcal{O}_w$.

\begin{table}[t]
  \caption{Comparison with existing RTL generation benchmarks. {\bench} scale is 98 unique designs $\times$ 4 HDLs ($=$392 instances).}
  \label{tab:bench-compare}
  \centering
  \footnotesize
  \setlength{\tabcolsep}{3.5pt}
  \begin{tabular}{@{}l r c c c c@{}}
    \toprule
    Benchmark & Scale & Black-box sec.\ TB & Manual review & Multi-HDL\\
    \midrule
    VerilogEval  & ${\sim}$156  & \ding{55}   & \ding{55}    & \ding{55}\\
    RTLLM        & 32           & \ding{55}   & \ding{51}    & \ding{55}\\
    HardSecBench & 924          & Partial     & Sampled      & \ding{55}\\
    SecV         & 23           & N/A         & \ding{51}    & \ding{55}\\
    SecFSM       & 25           & N/A         & \ding{51}    & \ding{55}\\
    \midrule
    \textbf{{\bench}} & \textbf{392} & \ding{51} & \ding{51} & \ding{51}\\
    \bottomrule
  \end{tabular}
\end{table}

\subsubsection{Quality Assurance}
\label{sec:bench-qa}

\textbf{Testbench Coverage:}
We measure statement coverage of each golden implementation under its combined functional and security testbenches.
Across all 98 designs, the average line coverage is 95.84\%.
This is a sanity check that the testbenches exercise most of the golden RTL; it does not by itself prove that every obligation in $\mathcal{O}_w$ is covered.
Obligation coverage is instead reviewed manually (below) against the scenarios for each CWE family.

\textbf{Manual Review:}
Every case is reviewed independently by three engineers, each with at least three years of industry experience.
Reviewers check that (i)~the specification is functionally clear and complete, while security obligations are intentionally withheld; (ii)~the golden implementation is correct and meets both functional and security requirements; and (iii)~the testbenches are sound and cover the relevant security obligations.
Conflicts are discussed until the three reviewers reach consensus.

\subsubsection{Comparison with Existing Benchmarks}
\label{sec:bench-compare}

Table~\ref{tab:bench-compare} compares {\bench} with prior work.
Functional benchmarks~\cite{liu2023verilogeval,lu2024rtllm} omit security; security-aware sets are either synthetic and single-HDL~\cite{chen2026hardsecbench} or small and unreleased (SecV~\cite{fan2025secv}, SecFSM~\cite{hu2026secfsm}).
{\bench} instead uses real SoC IP, full manual review, four HDLs, and black-box security testbenches (98 designs $\times$ 4 HDLs $=$ 392 instances).

\begin{table*}[t]
  \caption{Functional (F), security (S), and all-pass (A) rates (\%) on {\bench} (mean$\pm$std over 5 runs). L0: Vanilla; L1: Self-think; L2: CWE-Vanilla (\S\ref{sec:prompt-strategies}). Avg.\ rows are lightly highlighted. Bold marks the best mean per metric within each language row. A is the decisive metric.}
  \label{tab:prompting-results}
  \centering
  \renewcommand{\arraystretch}{1.05}
  \begin{tabular}{@{}ll *{3}{c} *{3}{c} *{3}{c}@{}}
    \toprule
    & & \multicolumn{3}{c}{Functional} & \multicolumn{3}{c}{Security} & \multicolumn{3}{c}{All-pass} \\
    \cmidrule(lr){3-5} \cmidrule(lr){6-8} \cmidrule(l){9-11}
    Model & Lang. & L0 & L1 & L2 & L0 & L1 & L2 & L0 & L1 & L2 \\
    \midrule
    % ----- DS-v4 -----
      & Verilog
        & \textbf{76.3}$\pm$1.9 & 56.5$\pm$2.6 & 56.9$\pm$1.2
        & 15.7$\pm$0.5 & 58.2$\pm$0.9 & \textbf{64.7}$\pm$2.6
        & 8.2$\pm$1.1 & 41.2$\pm$2.6 & \textbf{46.9}$\pm$2.2 \\
      & SystemVerilog
        & \textbf{81.6}$\pm$1.8 & 70.8$\pm$4.0 & 69.8$\pm$1.5
        & 14.7$\pm$1.0 & 64.3$\pm$4.7 & \textbf{68.8}$\pm$1.9
        & 9.8$\pm$0.5 & 52.4$\pm$5.3 & \textbf{59.8}$\pm$1.5 \\
      & VHDL
        & \textbf{63.3}$\pm$0.9 & 54.6$\pm$2.3 & 49.4$\pm$3.6
        & 9.6$\pm$1.9 & 48.9$\pm$3.3 & \textbf{49.2}$\pm$2.0
        & 8.0$\pm$1.6 & 42.2$\pm$2.9 & \textbf{45.1}$\pm$3.2 \\
      & Python
        & \textbf{72.5}$\pm$1.7 & 62.5$\pm$1.8 & 55.9$\pm$4.3
        & 18.0$\pm$1.7 & 56.8$\pm$3.1 & \textbf{63.1}$\pm$4.9
        & 7.5$\pm$1.7 & 44.1$\pm$1.2 & \textbf{48.0}$\pm$4.4 \\
      \rowcolor{blue!6}
      \multirow{-5}{*}{DS-v4}
      & Avg.
        & \textbf{73.4}$\pm$0.3 & 61.1$\pm$1.6 & 58.0$\pm$1.1
        & 14.5$\pm$0.8 & 57.0$\pm$1.0 & \textbf{61.4}$\pm$1.6
        & 8.4$\pm$0.6 & 45.0$\pm$1.0 & \textbf{50.0}$\pm$1.5 \\
    \midrule
    % ----- GLM-5.2 -----
      & Verilog
        & \textbf{84.7}$\pm$2.8 & 70.6$\pm$3.3 & 66.9$\pm$3.8
        & 21.2$\pm$1.9 & 51.6$\pm$4.2 & \textbf{64.3}$\pm$3.2
        & 16.9$\pm$2.1 & 44.3$\pm$3.3 & \textbf{54.1}$\pm$2.3 \\
      & SystemVerilog
        & \textbf{83.1}$\pm$1.4 & 69.0$\pm$2.4 & 60.8$\pm$2.9
        & 19.2$\pm$1.8 & 53.1$\pm$2.6 & \textbf{62.5}$\pm$1.6
        & 15.5$\pm$1.6 & 46.3$\pm$2.4 & \textbf{50.2}$\pm$3.7 \\
      & VHDL
        & \textbf{64.9}$\pm$1.7 & 48.6$\pm$1.9 & 47.5$\pm$1.4
        & 11.0$\pm$1.5 & 41.0$\pm$2.5 & \textbf{46.9}$\pm$3.3
        & 10.2$\pm$1.7 & 34.9$\pm$2.5 & \textbf{39.4}$\pm$2.0 \\
      & Python
        & \textbf{72.0}$\pm$3.1 & 59.6$\pm$3.3 & 49.6$\pm$3.3
        & 19.2$\pm$2.5 & 46.1$\pm$0.8 & \textbf{49.4}$\pm$3.2
        & 14.3$\pm$2.2 & 35.7$\pm$1.7 & \textbf{43.5}$\pm$3.3 \\
      \rowcolor{blue!6}
      \multirow{-5}{*}{GLM-5.2}
      & Avg.
        & \textbf{76.2}$\pm$1.1 & 61.9$\pm$1.4 & 56.2$\pm$1.8
        & 17.6$\pm$0.8 & 48.0$\pm$1.6 & \textbf{55.8}$\pm$1.3
        & 14.2$\pm$1.0 & 40.3$\pm$1.7 & \textbf{46.8}$\pm$1.6 \\
    \midrule
    % ----- GPT-5.6 -----
      & Verilog
        & \textbf{92.5}$\pm$0.8 & 78.4$\pm$3.8 & 70.2$\pm$0.8
        & 42.7$\pm$0.8 & 61.2$\pm$2.8 & \textbf{65.1}$\pm$1.6
        & 40.6$\pm$0.8 & 54.7$\pm$2.0 & \textbf{59.0}$\pm$1.8 \\
      & SystemVerilog
        & \textbf{85.9}$\pm$1.0 & 77.5$\pm$1.8 & 70.4$\pm$1.3
        & 47.1$\pm$1.8 & 62.3$\pm$1.3 & \textbf{69.0}$\pm$2.4
        & 40.8$\pm$1.3 & 54.1$\pm$2.1 & \textbf{59.6}$\pm$2.0 \\
      & VHDL
        & \textbf{65.5}$\pm$1.5 & 54.7$\pm$3.2 & 49.0$\pm$3.0
        & 24.1$\pm$2.1 & 43.9$\pm$1.6 & \textbf{48.4}$\pm$1.9
        & 22.4$\pm$2.1 & 38.4$\pm$1.8 & \textbf{44.3}$\pm$3.5 \\
      & Python
        & \textbf{69.0}$\pm$2.3 & 58.0$\pm$2.8 & 53.3$\pm$2.2
        & 27.6$\pm$1.3 & 49.4$\pm$3.1 & \textbf{56.1}$\pm$3.4
        & 22.2$\pm$1.5 & 40.4$\pm$2.6 & \textbf{45.9}$\pm$3.1 \\
      \rowcolor{blue!6}
      \multirow{-5}{*}{GPT-5.6}
      & Avg.
        & \textbf{78.2}$\pm$0.7 & 67.1$\pm$0.5 & 60.7$\pm$0.9
        & 35.4$\pm$0.7 & 54.2$\pm$0.8 & \textbf{59.6}$\pm$1.9
        & 31.5$\pm$0.9 & 46.9$\pm$0.4 & \textbf{52.2}$\pm$1.9 \\
    \midrule
    % ----- Mimo-v2.5 -----
      & Verilog
        & \textbf{76.7}$\pm$1.6 & 67.3$\pm$5.1 & 65.5$\pm$1.5
        & 38.6$\pm$3.8 & 61.0$\pm$2.7 & \textbf{66.9}$\pm$2.9
        & 33.7$\pm$3.5 & 53.5$\pm$3.4 & \textbf{55.5}$\pm$1.8 \\
      & SystemVerilog
        & \textbf{78.4}$\pm$2.2 & 68.3$\pm$1.8 & 66.1$\pm$3.6
        & 33.9$\pm$2.5 & 64.8$\pm$2.9 & \textbf{70.8}$\pm$1.5
        & 28.4$\pm$2.8 & 54.5$\pm$2.0 & \textbf{59.6}$\pm$2.7 \\
      & VHDL
        & \textbf{59.6}$\pm$1.5 & 53.1$\pm$1.7 & 49.4$\pm$2.0
        & 17.8$\pm$1.0 & 45.7$\pm$2.2 & \textbf{50.8}$\pm$1.6
        & 15.9$\pm$0.5 & 43.3$\pm$1.9 & \textbf{44.3}$\pm$1.9 \\
      & Python
        & \textbf{82.0}$\pm$1.5 & 70.8$\pm$2.2 & 67.3$\pm$2.7
        & 24.7$\pm$1.8 & 59.8$\pm$2.5 & \textbf{65.5}$\pm$3.4
        & 21.2$\pm$2.0 & 53.5$\pm$1.5 & \textbf{56.5}$\pm$2.4 \\
      \rowcolor{blue!6}
      \multirow{-5}{*}{Mimo-v2.5}
      & Avg.
        & \textbf{74.2}$\pm$0.9 & 64.9$\pm$1.3 & 62.1$\pm$1.7
        & 28.7$\pm$1.3 & 57.8$\pm$1.2 & \textbf{63.5}$\pm$1.2
        & 24.8$\pm$1.4 & 51.2$\pm$0.7 & \textbf{54.0}$\pm$1.8 \\
    \midrule
    % ----- MM-M3 -----
      & Verilog
        & \textbf{88.2}$\pm$1.0 & 73.7$\pm$5.4 & 74.7$\pm$4.4
        & 22.4$\pm$3.6 & 60.8$\pm$2.8 & \textbf{65.5}$\pm$4.9
        & 18.0$\pm$2.9 & 52.6$\pm$3.8 & \textbf{58.6}$\pm$4.5 \\
      & SystemVerilog
        & \textbf{85.3}$\pm$2.0 & 65.9$\pm$4.6 & 68.2$\pm$5.8
        & 23.1$\pm$3.0 & 59.7$\pm$5.1 & \textbf{62.9}$\pm$3.1
        & 19.0$\pm$2.5 & 48.8$\pm$3.5 & \textbf{51.6}$\pm$3.9 \\
      & VHDL
        & \textbf{67.2}$\pm$2.0 & 53.4$\pm$3.8 & 48.8$\pm$4.8
        & 18.2$\pm$3.4 & 44.8$\pm$2.4 & \textbf{46.7}$\pm$5.9
        & 15.5$\pm$3.3 & 37.4$\pm$1.7 & \textbf{39.2}$\pm$4.6 \\
      & Python
        & \textbf{76.9}$\pm$2.7 & 54.6$\pm$2.5 & 63.1$\pm$2.3
        & 16.5$\pm$3.0 & 50.9$\pm$4.4 & \textbf{52.0}$\pm$3.6
        & 12.0$\pm$2.1 & 40.2$\pm$1.9 & \textbf{42.9}$\pm$2.7 \\
      \rowcolor{blue!6}
      \multirow{-5}{*}{MM-M3}
      & Avg.
        & \textbf{79.4}$\pm$1.0 & 61.9$\pm$0.9 & 63.7$\pm$3.3
        & 20.1$\pm$1.5 & 54.1$\pm$1.3 & \textbf{56.8}$\pm$2.4
        & 16.1$\pm$1.2 & 44.7$\pm$1.2 & \textbf{48.1}$\pm$2.3 \\
    \bottomrule
  \end{tabular}
  \vspace{-0.5cm}
\end{table*}

\subsection{Empirical Setup}
\label{sec:empirical-setup}

\subsubsection{Models}
We evaluate five frontier LLMs spanning major providers: GPT-5.6-Luna, Mimo-v2.5-Pro, MiniMax-M3, GLM-5.2, and DeepSeek-v4-Pro.\footnote{All models are accessed through their official APIs in a zero-shot setting with temperature 0.6 and non-thinking mode.}
Each experiment is repeated five times per case per language; we report mean and standard deviation.
This section characterizes the security deficit rather than ranking methods, so we do not run pairwise significance tests.

\subsubsection{Prompting Strategies}
\label{sec:prompt-strategies}

We use three prompting strategies that vary the security knowledge given to $G_\theta$:

\begin{enumerate}[leftmargin=*,nosep]
  \item \textbf{Vanilla (L0).} The model receives only the functional specification $s$ and draws code $c \sim G_\theta(s)$. No security information is provided.
  \item \textbf{Self-think (L1).} Before generating code, the model must first write a security analysis of $s$, then generate RTL. This tests whether $G_\theta$ can infer obligations on its own.
  \item \textbf{CWE-Vanilla (L2).} The prompt includes the CWE entry aligned with the case, using the official identifier and description (e.g., ``CWE-226: Sensitive Information in Resource Not Removed Before Reuse''). The model is asked to analyze the security obligations by the given CWE entry and then generate RTL.
\end{enumerate}

L0 is the no-knowledge baseline.
L1 keeps the setting closed: the model may reason, but receives no CWE knowledge.
L2 is an \textit{intentional} upper-bound cue: it assumes the relevant hardware weakness is known, as in CWE-retrieval pipelines.
L2 still leaves concrete obligations implicit; the model must turn the CWE entry into design-specific constraints by itself.

\subsubsection{Evaluation Metrics}
\label{sec:emp-metrics}

We score each generated module $c$ with three binary indicators defined by the black-box testbenches in \S\ref{sec:bench-tb}:

\begin{itemize}[leftmargin=*,nosep]
  \item \textbf{Functional pass (F)}: $c \models T_\mathit{func}$.
  \item \textbf{Security pass (S)}: $c \models T_\mathit{sec}$.
  \item \textbf{All-pass (A)}: $c \models T_\mathit{func} \wedge c \models T_\mathit{sec}$.
\end{itemize}

Rates are aggregated over cases and languages and reported as mean$\pm$std across five runs.
All-pass is the decisive metric: raising F or S alone is insufficient when the other fails.
Compile or simulation failures count as neither F nor S.
Language-specific simulators are Icarus Verilog\footnote{\url{https://github.com/steveicarus/iverilog}} (\texttt{-g2012}) for Verilog and SystemVerilog, GHDL\footnote{\url{https://github.com/ghdl/ghdl}} for VHDL, and Amaranth's built-in simulator\footnote{\url{https://github.com/amaranth-lang/amaranth}} for Python.

\subsection{Empirical Results}
\label{sec:empirical-results}

Table~\ref{tab:prompting-results} reports F, S, and A across the three prompting strategies and five models.
We organize the analysis around two findings.

\subsubsection{Finding 1: CWE Knowledge Helps, but Obligations Stay Implicit}

Under the standard baseline (L0), all five models achieve high average functional pass rates (73.4\%--79.4\%), yet their security pass rates remain strikingly low (14.5\%--35.4\%).
This functional-security disparity persists across all target languages. Crucially, strong functional capability does not guarantee security; for instance, MiniMax-M3 achieves the highest average functional rate (79.4\%) but ranks near the bottom in security performance (20.1\%).

Providing aligned CWE knowledge (L2) substantially mitigates this gap, elevating mean security from 23.3\% to 59.4\% ($L0 < L2$ holds strictly across all models and languages). 
This marked improvement confirms that low security under L0 stems not from an inability to synthesize defensive RTL, but from a lack of weakness awareness during generation.

\begin{tcolorbox}[colback=gray!5, colframe=gray!60, boxrule=0.4pt, left=4pt, right=4pt, top=2pt, bottom=2pt]
\textbf{Finding 1.}
Standard specifications (L0) yield high functional correctness (${\sim}$76\%) but severe security vulnerabilities (14\,--\,35\%). Providing explicit CWE knowledge (L2) raises mean security to 59\%, proving that weakness awareness—rather than RTL synthesis capability—is the primary bottleneck.
\end{tcolorbox}

\subsubsection{Finding 2: Self-Thinking Helps, but Cannot Replace CWE Knowledge}

Prompting models to infer security obligations autonomously (L1) yields substantial gains over L0, boosting mean security from 23.3\% to 54.2\% and all-pass from 19.0\% to 45.6\%. This $L1 > L0$ shift proves that LLMs possess latent defensive capabilities that can be activated through prompt-induced self-thinking.

However, L1 consistently lags behind L2 across all models (mean S: 54.2\% vs.\ 59.4\%; mean A: 45.6\% vs.\ 50.2\%): self-thinking alone cannot replace explicit CWE knowledge. Because hardware specifications frequently leave security obligations implicit, unassisted self-thinking cannot fully bridge the domain knowledge gap without explicit CWE guidance.
Furthermore, both L1 and L2 induce a clear functional penalty: mean functional pass rate drops from 76.3\% (L0) to 63.4\% (L1) and 60.1\% (L2). 

\begin{tcolorbox}[colback=gray!5, colframe=gray!60, boxrule=0.4pt, left=4pt, right=4pt, top=2pt, bottom=2pt]
\textbf{Finding 2.}
Self-thinking ($L1$) effectively activates latent defenses ($L1 > L0$), but cannot replace explicit CWE knowledge ($L1 < L2$) due to implicit spec gaps. Moreover, these methods trade functional correctness for security ($F$ drops from 76\% to ${\sim}$60\%), capping overall all-pass at roughly 50\%.
\end{tcolorbox}

\subsection{Implications}
\label{sec:empirical-implications}

Findings~1--2 motivate \tool{}.
CWE knowledge is needed, self-thinking cannot replace it, and single-pass security prompting still trades functionality for security.
The remaining task is to \textit{infer} signal-level obligations from the specification, then apply them without degrading functional correctness.

\section{\tool{}: Inferring Implicit Security Obligations}
\label{sec:method}

Following \S\ref{sec:empirical}, \tool{} infers implicit security obligations rather than prompting the LLM to invent them.
We treat many such obligations as \textit{mitigation-evidence gaps}: $s$ shows weakness preconditions (e.g., a reusable secret register is written) but not the required mitigation (e.g., a clear at done).
Matching $s$ against a reusable mitigation ontology then reduces inference to a deterministic signal-level gap check.

\subsection{Overview and Design Goals}
\label{sec:overview}

These observations yield three design goals for \tool{}:
\begin{itemize}[leftmargin=*,nosep]
  \item \textbf{G1 (Inferable).}
 Obligations are inferred from structural evidence in the specification together with an explicit security ontology, not guessed end-to-end by an LLM.
  \item \textbf{G2 (Auditable).}
 Each inferred obligation is traceable to matched structures, uncovered elements, and a CWE scenario, forming a reproducible witness rather than an opaque model belief.
  \item \textbf{G3 (Functionality-preserving).}
 Patch the design to satisfy the inferred security obligations while changing its functional behavior as little as possible.
\end{itemize}

\begin{figure*}[!t]
  \centering
  \includegraphics[width=0.92\linewidth]{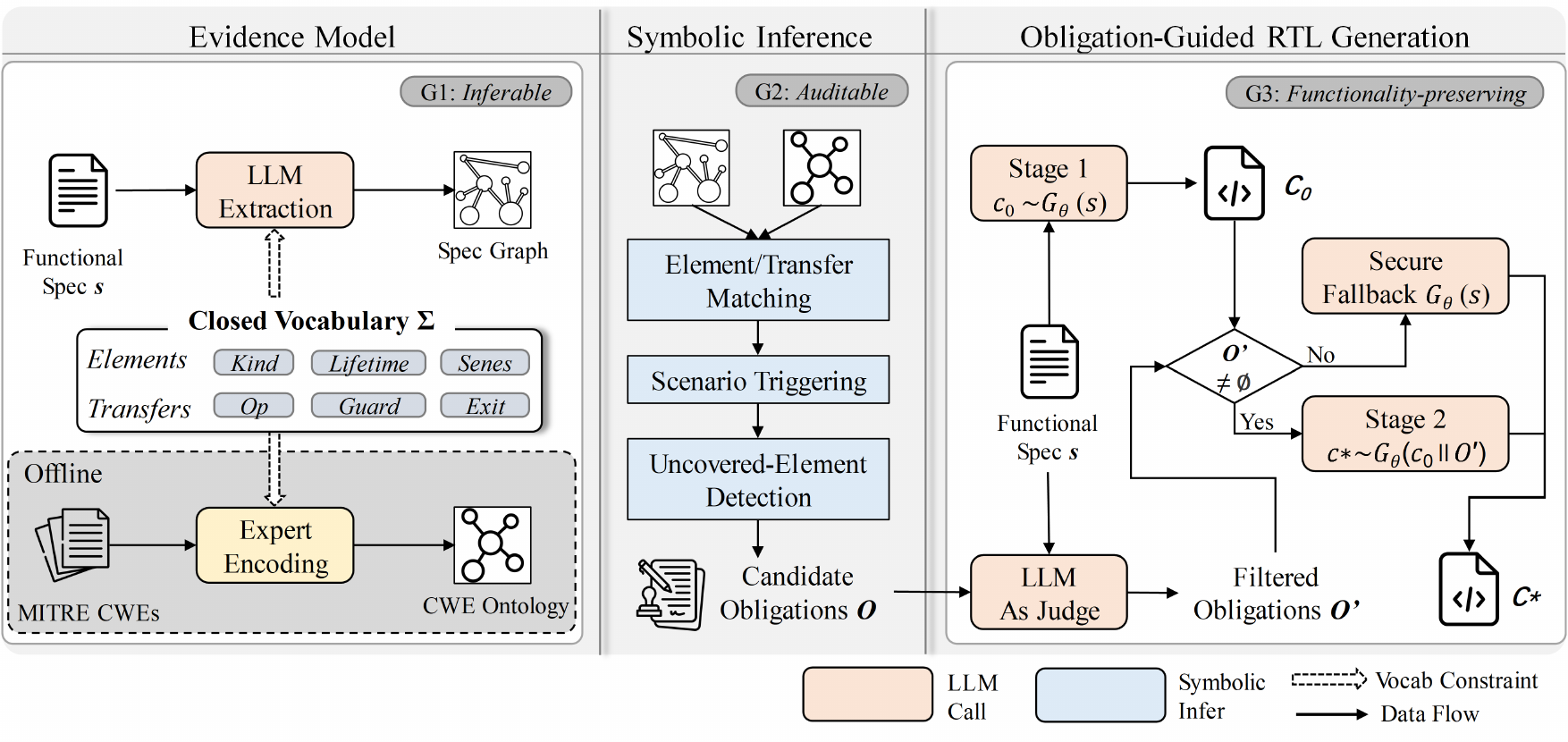}
  \caption{Overview of the \tool{} pipeline.}
  \label{fig:pipeline}
\end{figure*}

\tool{} realizes these goals through a three-stage pipeline (Figure~\ref{fig:pipeline}).
An \textbf{evidence model} binds a per-specification \textit{Functional-Semantic Graph} (FSG) to a reusable CWE pattern ontology over a shared closed vocabulary~(\S\ref{sec:evidence}).
A \textbf{symbolic engine} infers signal-level \textit{security obligations} by checking for mitigation-evidence gaps and pinpointing the responsible elements, with all gap decisions made symbolically~(\S\ref{sec:symbolic-infer}).
Finally, an LLM filters inapplicable obligations and generates RTL via obligation-guided revision~(\S\ref{sec:codegen}).

\subsection{Evidence Model}
\label{sec:evidence}

Inferring implicit obligations requires two complementary kinds of structured evidence: \textit{what security-relevant structure is present in a given specification?} and \textit{what mitigation evidence should be present for that structure?}
\tool{} encodes both as graphs over a shared closed vocabulary~$\Sigma$, so that the symbolic engine can answer these questions by matching.

\subsubsection{Closed Vocabulary $\Sigma$}
\label{sec:sigma}

All structural labels in \tool{} are drawn from a fixed vocabulary:
\begin{equation}\label{eq:sigma}
  \Sigma = (\mathit{Kind},\; \mathit{Lifetime},\; \mathit{Sens},\; \mathit{Op},\; \mathit{Guard},\; \mathit{Exit})
\end{equation}
where each dimension is a finite set (Table~\ref{tab:sigma}).

The first three dimensions characterize \emph{elements}:
$\mathit{Kind}$ denotes the hardware type (e.g., register, buffer),
$\mathit{Lifetime}$ indicates how long the element's data persists (e.g., reusable across transactions vs.\ local to one),
and $\mathit{Sens}$ records its security sensitivity level.

The remaining three characterize \emph{transfers}:
$\mathit{Op}$ specifies the operation performed (e.g., clear, latch, crypto),
$\mathit{Guard}$ captures the control condition under which the transfer fires (e.g., on reset or state transition),
and $\mathit{Exit}$ identifies the control-state phase in which the transfer occurs (e.g., done, idle, error).

\begin{table}[h]
  \caption{Closed vocabulary $\Sigma$. Every element and transfer label is drawn exclusively from these six finite sets.}
  \label{tab:sigma}
  \centering
  \small
  \begin{tabular}{@{}l>{\raggedright\arraybackslash}p{0.8\columnwidth}@{}}
    \toprule
    Dimension & Values \\
    \midrule
    $\mathit{Kind}$     & \texttt{port, register, memory, buffer, bus} \\
    $\mathit{Lifetime}$ & \texttt{reusable, local, parameter} \\
    $\mathit{Sens}$     & \texttt{secret, internal, public} \\
    $\mathit{Op}$       & \texttt{clear, latch, mux, validate, configure, crypto, debug} \\
    $\mathit{Guard}$    & \texttt{reset, transition, error} \\
    $\mathit{Exit}$     & \texttt{idle, success, done, reset, error} \\
    \bottomrule
  \end{tabular}
\end{table}

Because every label comes from a finite set, the matching predicates in \S\ref{sec:matching} always terminate.
Deterministic evaluation then makes every structural decision reproducible.

\subsubsection{Spec Graph $G_\mathit{spec}$}
\label{sec:gspec}

The functional-semantic graph (FSG) is the specification-side evidence extracted from one functional specification:
\begin{equation}\label{eq:fsg}
  G_\mathit{spec} = (E, T, S)
\end{equation}
\begin{itemize}[leftmargin=*,nosep]
  \item $E \subseteq \mathit{Name} \times \mathit{Kind} \times \mathit{Lifetime} \times \mathit{Sens}$: stateful elements.
    Example: \texttt{(key\_reg, register, reusable, secret)}.
  \item $T \subseteq \mathit{Id} \times \mathit{Op} \times E \times \mathit{Guard}^* \times (S \cup \{\varepsilon\})$: transfers (operation, target, guards, and exit phase; $\varepsilon$ if unbound).
  \item $S \subseteq \mathit{Exit}$: control-state phases of the module lifecycle (e.g., \texttt{done}, \texttt{idle}, \texttt{error}).
\end{itemize}

A well-formedness constraint requires $\forall\, t \in T{:}\; t.\mathit{exit} \in S \cup \{\varepsilon\}$.
$G_\mathit{spec}$ records structural evidence for later inference; it does not decide whether any obligation is required.

The LLM translates the unstructured functional specification into $G_\mathit{spec}$.
It receives the specification text together with~$\Sigma$ as a schema constraint and outputs a JSON object conforming to the FSG schema.

\subsubsection{CWE Pattern Ontology $G_\mathit{onto}$}
\label{sec:gonto}

While $G_\mathit{spec}$ describes one design, $G_\mathit{onto}$ encodes reusable security knowledge: for a given structure, what mitigation evidence should be present.
$G_\mathit{onto}$ is built from the MITRE CWE definitions for the resource-access families in this paper.
Each CWE entry is a JSON file over~$\Sigma$.

For each family $w$, $G_\mathit{onto}$ contains:
\begin{itemize}[leftmargin=*,nosep]
  \item \textbf{Element patterns} $P_E$: constraints on kind, lifetime, and sensitivity (e.g., a \texttt{reusable}/\texttt{secret} element).
  \item \textbf{Transfer patterns} $P_T$: constraints on operation, guards, exit, and target (e.g., a \texttt{clear} at \texttt{done} on the matched element).
  \item \textbf{Scenarios}: rules that list required element/transfer patterns, optional context transfers, and \textit{missing-transfer} patterns (the expected mitigation).
  \item \textbf{Hints}: natural-language obligation templates filled with specific signal names when a scenario fires.
\end{itemize}

$G_\mathit{spec}$ records what one specification contains; $G_\mathit{onto}$ states what complete mitigation evidence should look like.
Shared vocabulary~$\Sigma$ lets \S\ref{sec:symbolic-infer} connect the two by deterministic matching (G1--G2).

\subsection{Symbolic Inference}
\label{sec:symbolic-infer}

The symbolic core takes $G_\mathit{spec}$ and $G_\mathit{onto}$ and outputs signal-level \textit{security obligations}: instantiations of the abstract classes $\mathcal{O}_w$ (\S\ref{sec:bg-cwe}) for the current specification.
This stage makes no LLM calls; every decision is deterministic.

A scenario $\mathit{sc}$ fires when its structural preconditions match \textit{and} at least one written element still lacks the expected mitigation:
\begin{equation}\label{eq:scenario}
  \mathit{triggered}(\mathit{sc}, G_\mathit{spec}) \;\iff\;
    \mathit{reqE} \;\wedge\; \mathit{reqT} \;\wedge\; \mathit{ctxT} \;\wedge\; \mathit{missT}
\end{equation}
where:
\begin{itemize}[leftmargin=*,nosep]
  \item $\mathit{reqE}$: every required element pattern has $\geq 1$ match in $E$.
  \item $\mathit{reqT}$: every required transfer pattern has $\geq 1$ match in $T$ (e.g., a secret element is written).
  \item $\mathit{ctxT}$: every contextual transfer pattern has $\geq 1$ match.
  \item $\mathit{missT}$: the uncovered-element set $U$ is non-empty.
    Here $U$ is computed from $\mathit{sc}$'s \textit{missing-transfer} patterns (the mitigation that should be present): a written element is uncovered if it is targeted by a required transfer but not covered by any matching mitigation transfer.
\end{itemize}
Thus $\mathit{missT}$ means a mitigation-evidence gap, not the presence of mitigation.
This matches how implicit obligations appear in functional specifications: the weakness preconditions are present, while the required mitigation is absent.

\subsubsection{Matching and Inference}
\label{sec:matching}

Element matching checks whether an instance element $e \in E$ satisfies a pattern $p \in P_E$:
\begin{equation}\label{eq:elem-match}
  \mathit{match}(e, p) \;\iff\;
    \bigwedge_{d \in \{\mathit{Kind}, \mathit{Lifetime}, \mathit{Sens}\}}
      \bigl(p.d = \varnothing \;\vee\; e.d \in p.d\bigr)
\end{equation}
An empty constraint set is a wildcard.

Transfer matching also checks operation, guard tags, exit path, and target binding:
\begin{equation}\label{eq:trans-match}
\begin{split}
  \mathit{match}(t, p, \sigma) \;\iff\;&\;
    (p.\mathit{ops} = \varnothing \;\vee\; t.\mathit{op} \in p.\mathit{ops}) \\
    \wedge\;&\; (p.\mathit{target} = \varnothing \;\vee\; t.\mathit{target} \in \sigma(p.\mathit{target})) \\
    \wedge\;&\; p.\mathit{guard}^{+} \subseteq \mathit{tags}(t) \\
    \wedge\;&\; p.\mathit{guard}^{-} \cap \mathit{tags}(t) = \varnothing \\
    \wedge\;&\; (p.\mathit{exit} = \varnothing \;\vee\; t.\mathit{exit} \in p.\mathit{exit})
\end{split}
\end{equation}
where $\sigma$ maps element-pattern identifiers to matched instances, and $\mathit{tags}(t)$ is the set of $\mathit{Guard}$ tokens for $t$ after the keyword normalization above.
Matches are collected per pattern rather than per scenario.
The symbolic stage is thus high-recall by construction, and joint consistency is resolved by the downstream LLM-as-Judge (\S\ref{sec:obl-filter}), which sees the specification and rejects obligations that do not apply to the design.

\begin{algorithm}[t]
\caption{Symbolic Obligation Inference}\label{alg:trigger}
\KwIn{$G_\mathit{spec}$, $G_\mathit{onto}$\; }
\KwOut{Triggered scenarios with uncovered elements\; }
$\sigma_E \gets \{p.\mathit{id} : \{e \in E \mid \mathit{match}(e, p)\} \;\;\forall\, p \in P_E\}$\;
$\sigma_T \gets \{p.\mathit{id} : \{t \in T \mid \mathit{match}(t, p, \sigma_E)\} \;\;\forall\, p \in P_T\}$\;
$R \gets \varnothing$\;
\ForEach{scenario $\mathit{sc} \in G_\mathit{onto}.\mathit{scenarios}$}{
  $\mathit{reqE\_ok} \gets \bigwedge_{p \in \mathit{sc}.\mathit{req\_elems}} \sigma_E[p] \neq \varnothing$\;
  $\mathit{reqT\_ok} \gets \bigwedge_{p \in \mathit{sc}.\mathit{req\_trans}} \sigma_T[p] \neq \varnothing$\;
  $\mathit{ctx\_ok} \gets \bigwedge_{p \in \mathit{sc}.\mathit{ctx\_trans}} \sigma_T[p] \neq \varnothing$\;
  $U \gets \textsc{UncoveredElements}(G_\mathit{spec}, \mathit{sc}, \sigma_T)$\;
  \If{$\mathit{reqE\_ok} \,\wedge\, \mathit{reqT\_ok} \,\wedge\, \mathit{ctx\_ok} \,\wedge\, U \neq \varnothing$}{
    $R \gets R \cup \{(\mathit{sc},\, \sigma_E,\, \sigma_T,\, U)\}$\;
  }
}
\Return $R$\;
\end{algorithm}

Algorithm~\ref{alg:trigger} evaluates every scenario in $G_\mathit{onto}$ against $G_\mathit{spec}$.
The last conjunct is exactly $\mathit{missT}$ in Equation~\ref{eq:scenario}: preconditions hold and $U \neq \varnothing$.

\subsubsection{Uncovered-Element Detection}
\label{sec:uncovered}

Let $E_w \subseteq E$ be the \textit{written elements}: elements in the FSG targeted by the scenario's required transfers.
\textsc{UncoveredElements} compares $E_w$ to the scenario's missing-transfer patterns (expected mitigations).

For each missing-transfer pattern $m$:
\begin{itemize}[leftmargin=*,nosep]
  \item \textbf{Per-element check.} If $m$ binds to a target element, written elements already covered by a matching mitigation transfer are dropped; the rest join $U$: $U \gets U \cup (E_w \setminus \mathit{mitigated}_m)$.
  \item \textbf{Global check.} If $m$ has no target binding, and no transfer in the design matches $m$, then every element in $E_w$ is uncovered.
\end{itemize}

This yields signal-level output (G2).
For example, instead of ``beware CWE-226,'' \tool{} can report that \texttt{plaintext\_reg} and \texttt{key\_reg} lack a \texttt{clear} at \texttt{done}.

\subsubsection{Obligation Instantiation}
\label{sec:obl-inst}

Each triggered scenario has one or more hint templates in $G_\mathit{onto}$.
An inferred \textit{security obligation} is:
\begin{equation}\label{eq:obl}
  o = (\mathit{sc},\; \mathit{cwe},\; U,\; \mathit{text})
\end{equation}
where $\mathit{sc}$ is the scenario, $\mathit{cwe}$ is the CWE id, $U$ is the uncovered-element set, and $\mathit{text}$ is the filled natural-language statement.

Placeholders are filled as:
\begin{itemize}[leftmargin=*,nosep]
  \item \texttt{\{elem\_names\}} $\gets U$;
  \item \texttt{\{trans\_names\}} $\gets$ operations targeting elements in $U$;
  \item \texttt{\{exit\_hints\}} $\gets$ exit keywords from the scenario.
\end{itemize}
The pair $(\mathit{sc}, U)$ is the structural witness for $o$.
Let $O = \{o_1, \dots, o_n\}$ be all candidate obligations for the specification.
$O$ is high-recall: every structurally supported gap is kept for the next stage.

\subsection{Obligation-Guided RTL Generation}
\label{sec:codegen}

The final stage filters $O$ and generates RTL while changing functional behavior as little as possible (G3).

\subsubsection{Obligation Filtering via LLM-as-Judge}
\label{sec:obl-filter}

Some candidates are structural false positives.
An \textit{LLM-as-Judge} re-checks each $o_i$ against $s$ and an FSG summary:
\begin{equation}\label{eq:filter}
  \mathit{filter}: O \times s \times G_\mathit{spec} \;\to\; O' \subseteq O
\end{equation}
The judge may only keep or reject; it cannot invent obligations.
If its output cannot be parsed, we set $O'{=}O$ so filtering never silently drops a true candidate.

\subsubsection{Functionality-Preserving Two-Stage Generation}
\label{sec:rtl-gen}

One-shot generation $c \sim G_\theta(s \,\|\, O')$ recreates the functional penalty in Finding~2.
We therefore use two stages.
Stage~1 builds a functional draft from the specification alone:
\begin{equation}\label{eq:stage1}
  c_0 \sim G_\theta(s)
\end{equation}
Stage~2 revises $c_0$.
If $O' \neq \varnothing$, revision is obligation-guided:
\begin{equation}\label{eq:stage2}
  c^* \sim G_\theta(c_0 \,\|\, O')
\end{equation}
Because obligations name uncovered elements in $U$, edits are local patches rather than a full rewrite.
If $O' = \varnothing$ or FSG extraction fails, Stage~2 uses a \textit{secure fallback} $c^* \sim G_\theta(c_0 \,\|\, s_{\mathit{fb}})$, asking the model to harden $c_0$ while preserving function and the module interface.
This keeps the pipeline uniform when the symbolic path yields no obligations.

\section{Results Analysis}
\label{sec:results}

\begin{table*}[t]
\caption{Functional, security, and all-pass rates (\%) on {\bench} (mean$\pm$std over 5 runs).
Avg.\ rows are lightly highlighted.
Bold marks the best mean per metric within each language row.}
\label{tab:tool-results}
\centering
\begin{tabular}{@{}ll *{3}{c} *{3}{c} *{3}{c}@{}}
  \toprule
  & & \multicolumn{3}{c}{Functional} & \multicolumn{3}{c}{Security} & \multicolumn{3}{c}{All-pass} \\
  \cmidrule(lr){3-5} \cmidrule(lr){6-8} \cmidrule(l){9-11}
  Model & Lang. & SecV & RESCUE & {\tool} & SecV & RESCUE & {\tool} & SecV & RESCUE & {\tool} \\
  \midrule
  % ----- DS-v4 -----
    & Verilog
      & 58.0$\pm$5.1 & 53.1$\pm$3.0 & \textbf{66.1}$\pm$1.8
      & 67.3$\pm$2.8 & \textbf{70.4}$\pm$3.6 & 68.8$\pm$3.5
      & 48.6$\pm$4.5 & 48.2$\pm$1.6 & \textbf{55.1}$\pm$3.3 \\
    & SV
      & 62.3$\pm$1.3 & 70.4$\pm$2.8 & \textbf{76.3}$\pm$2.1
      & 72.9$\pm$1.8 & \textbf{81.0}$\pm$3.4 & 78.4$\pm$1.0
      & 55.7$\pm$0.5 & 65.5$\pm$4.6 & \textbf{66.3}$\pm$1.6 \\
    & VHDL
      & 47.5$\pm$3.1 & 51.0$\pm$1.8 & \textbf{58.2}$\pm$2.3
      & 51.2$\pm$2.0 & 53.3$\pm$1.5 & \textbf{55.9}$\pm$1.2
      & 44.5$\pm$2.6 & 49.0$\pm$1.8 & \textbf{49.6}$\pm$2.4 \\
    & Python
      & 53.7$\pm$2.4 & 55.9$\pm$1.8 & \textbf{67.3}$\pm$2.6
      & 64.5$\pm$2.7 & 69.0$\pm$2.9 & \textbf{76.1}$\pm$2.3
      & 48.0$\pm$2.3 & 50.6$\pm$1.4 & \textbf{60.0}$\pm$3.3 \\
    \rowcolor{blue!6}
    \multirow{-5}{*}{DS-v4}
    & Avg.
      & 55.4$\pm$0.5 & 57.6$\pm$1.5 & \textbf{67.0}$\pm$0.9
      & 64.0$\pm$1.6 & 68.4$\pm$1.1 & \textbf{69.8}$\pm$1.6
      & 49.2$\pm$0.6 & 53.3$\pm$1.4 & \textbf{57.8}$\pm$1.8 \\
  \midrule
  % ----- GLM-5.2 -----
    & Verilog
      & 65.3$\pm$3.6 & 56.3$\pm$3.7 & \textbf{81.6}$\pm$3.8
      & 66.1$\pm$3.2 & 66.7$\pm$2.2 & \textbf{77.8}$\pm$3.1
      & 57.8$\pm$2.6 & 49.6$\pm$2.9 & \textbf{72.0}$\pm$4.9 \\
    & SV
      & 59.2$\pm$0.7 & 51.6$\pm$2.9 & \textbf{79.4}$\pm$2.3
      & 65.3$\pm$2.4 & 66.9$\pm$1.9 & \textbf{75.5}$\pm$2.1
      & 53.9$\pm$1.4 & 44.9$\pm$2.6 & \textbf{67.3}$\pm$1.4 \\
    & VHDL
      & 43.1$\pm$1.6 & 48.0$\pm$1.6 & \textbf{62.9}$\pm$1.7
      & 44.9$\pm$1.8 & 49.6$\pm$3.5 & \textbf{56.5}$\pm$0.5
      & 37.5$\pm$2.0 & 41.4$\pm$1.9 & \textbf{53.5}$\pm$0.5 \\
    & Python
      & 43.9$\pm$3.8 & 49.6$\pm$5.7 & \textbf{66.9}$\pm$1.0
      & 52.3$\pm$3.3 & 56.1$\pm$5.1 & \textbf{65.9}$\pm$3.5
      & 39.0$\pm$5.2 & 43.9$\pm$5.3 & \textbf{59.2}$\pm$2.2 \\
    \rowcolor{blue!6}
    \multirow{-5}{*}{GLM-5.2}
    & Avg.
      & 52.9$\pm$1.7 & 51.4$\pm$0.8 & \textbf{72.7}$\pm$1.2
      & 57.2$\pm$1.9 & 59.9$\pm$1.4 & \textbf{68.9}$\pm$1.5
      & 47.0$\pm$1.5 & 45.0$\pm$1.2 & \textbf{63.0}$\pm$1.1 \\
  \midrule
    % ----- GPT-5.6 -----
    & Verilog
    & 69.8$\pm$1.0 & 69.2$\pm$1.0 & \textbf{86.5}$\pm$1.0
    & 70.6$\pm$1.4 & 71.0$\pm$2.4 & \textbf{76.7}$\pm$1.2
    & 64.5$\pm$2.1 & 61.6$\pm$1.8 & \textbf{71.8}$\pm$1.0 \\
  & SV
    & 67.2$\pm$1.0 & 68.8$\pm$2.4 & \textbf{89.8}$\pm$1.4
    & 73.1$\pm$1.4 & 70.4$\pm$3.0 & \textbf{78.0}$\pm$0.8
    & 62.2$\pm$0.9 & 60.4$\pm$3.1 & \textbf{73.1}$\pm$1.2 \\
  & VHDL
    & 46.1$\pm$1.6 & 49.2$\pm$1.2 & \textbf{68.4}$\pm$1.7
    & 47.5$\pm$1.8 & 51.4$\pm$1.5 & \textbf{59.4}$\pm$1.0
    & 42.2$\pm$2.2 & 45.7$\pm$1.8 & \textbf{54.9}$\pm$2.1 \\
  & Python
    & 53.1$\pm$1.3 & 52.6$\pm$3.2 & \textbf{77.5}$\pm$2.2
    & 62.0$\pm$3.0 & 60.2$\pm$1.7 & \textbf{70.6}$\pm$1.4
    & 49.4$\pm$1.4 & 48.2$\pm$3.4 & \textbf{64.7}$\pm$2.2 \\
  \rowcolor{blue!6}
  \multirow{-5}{*}{GPT-5.6}
  & Avg.
    & 59.0$\pm$0.7 & 60.0$\pm$0.6 & \textbf{80.6}$\pm$0.2
    & 63.3$\pm$1.2 & 63.3$\pm$1.2 & \textbf{71.2}$\pm$0.5
    & 54.6$\pm$0.7 & 54.0$\pm$0.8 & \textbf{66.1}$\pm$0.4 \\
  \midrule
  % ----- Mimo-v2.5 -----
    & Verilog
      & 59.6$\pm$4.0 & 63.1$\pm$3.1 & \textbf{77.8}$\pm$2.4
      & 66.1$\pm$3.6 & 71.2$\pm$1.2 & \textbf{79.0}$\pm$2.9
      & 52.6$\pm$3.1 & 58.2$\pm$1.7 & \textbf{69.6}$\pm$3.3 \\
    & SV
      & 58.4$\pm$2.2 & 65.7$\pm$3.1 & \textbf{76.1}$\pm$2.9
      & 66.5$\pm$4.0 & 74.7$\pm$2.4 & \textbf{77.5}$\pm$1.4
      & 53.1$\pm$3.9 & 61.0$\pm$2.5 & \textbf{67.3}$\pm$2.7 \\
    & VHDL
      & 32.0$\pm$4.2 & 45.9$\pm$2.7 & \textbf{58.4}$\pm$2.0
      & 35.1$\pm$2.5 & 49.2$\pm$2.1 & \textbf{54.5}$\pm$2.1
      & 28.6$\pm$3.6 & 42.7$\pm$1.8 & \textbf{49.0}$\pm$1.4 \\
    & Python
      & 63.7$\pm$1.7 & 63.5$\pm$1.8 & \textbf{69.6}$\pm$2.6
      & 66.1$\pm$2.4 & 66.1$\pm$4.8 & \textbf{71.6}$\pm$3.8
      & 58.2$\pm$2.1 & 57.1$\pm$2.5 & \textbf{62.0}$\pm$1.8 \\
    \rowcolor{blue!6}
    \multirow{-5}{*}{Mimo-v2.5}
    & Avg.
      & 53.4$\pm$1.8 & 59.5$\pm$1.1 & \textbf{70.5}$\pm$1.2
      & 58.5$\pm$2.6 & 65.3$\pm$0.7 & \textbf{70.7}$\pm$1.6
      & 48.1$\pm$2.4 & 54.7$\pm$0.9 & \textbf{62.0}$\pm$1.4 \\
  \midrule
  % ----- MM-M3 -----
    & Verilog
      & 65.9$\pm$2.7 & 64.9$\pm$3.3 & \textbf{76.3}$\pm$4.6
      & 70.8$\pm$2.8 & 68.2$\pm$1.4 & \textbf{74.9}$\pm$1.7
      & 59.0$\pm$2.4 & 56.3$\pm$1.8 & \textbf{65.5}$\pm$4.0 \\
    & SV
      & 57.6$\pm$4.4 & 64.5$\pm$4.6 & \textbf{75.5}$\pm$1.7
      & 65.3$\pm$1.9 & 71.4$\pm$3.4 & \textbf{71.6}$\pm$2.5
      & 51.0$\pm$3.2 & 57.1$\pm$3.4 & \textbf{63.7}$\pm$3.5 \\
    & VHDL
      & 40.8$\pm$3.6 & 44.1$\pm$2.5 & \textbf{55.3}$\pm$4.0
      & 44.7$\pm$4.2 & 50.4$\pm$3.8 & \textbf{52.2}$\pm$4.9
      & 36.9$\pm$2.4 & 40.8$\pm$3.1 & \textbf{48.0}$\pm$4.6 \\
    & Python
      & 52.4$\pm$5.3 & 50.4$\pm$3.3 & \textbf{67.8}$\pm$2.4
      & 58.6$\pm$4.8 & 59.2$\pm$3.4 & \textbf{68.0}$\pm$1.9
      & 49.4$\pm$4.6 & 46.1$\pm$4.0 & \textbf{59.0}$\pm$2.1 \\
    \rowcolor{blue!6}
    \multirow{-5}{*}{MM-M3}
    & Avg.
      & 54.2$\pm$1.6 & 56.0$\pm$1.6 & \textbf{68.7}$\pm$1.3
      & 59.8$\pm$2.2 & 62.3$\pm$1.0 & \textbf{66.7}$\pm$1.5
      & 49.1$\pm$1.4 & 50.1$\pm$1.7 & \textbf{59.0}$\pm$2.1 \\
  \bottomrule
\end{tabular}
\end{table*}

\subsection{RQ1: How Does \tool{} Perform on Secure RTL Generation?}
\label{sec:rq1-baselines}

\subsubsection{Baselines}

To evaluate the effectiveness of \tool{} for secure RTL generation, we compare it with SecV~\cite{fan2025secv} and RESCUE~\cite{shi2026rescue}, two methods that add external security knowledge to LLM code generation.

\begin{itemize}
    \item \textbf{SecV} constructs a Hardware-CWE knowledge graph and uses clue-guided exploration to retrieve security subgraphs for secure Verilog generation.
    \item \textbf{RESCUE} is a RAG method for secure software generation: it stores CWE guidelines and sliced secure code examples, then retrieves them with multi-faceted search.
\end{itemize}

Both baselines obtain security guidance through the LLM by exploring a graph (SecV) or retrieving text and code (RESCUE).
\tool{} instead infers signal-level obligations with a symbolic engine~(\S\ref{sec:symbolic-infer}), filters them with an LLM-as-Judge~(\S\ref{sec:obl-filter}), and generates RTL in two stages: a functional draft, then an obligation-guided revision~(\S\ref{sec:rtl-gen}).
Security guidance thus comes from deterministic matching on the specification, not from free-form retrieved or generated advice.

\subsubsection{Setup}
Neither SecV nor RESCUE provides an artifact for {\bench}.
We re-implement both from their published designs and adapt them to our setting: a Hardware-CWE knowledge graph for SecV, and a knowledge base of CWE guidelines plus sliced secure code examples for RESCUE, each covering the five CWE families and four HDLs in the benchmark.
We then evaluate all three methods with the same five LLMs on {\bench}.
Each configuration is run five times; Table~\ref{tab:tool-results} reports mean$\pm$std for functional, security, and all-pass rates.

\subsubsection{Results}
\label{sec:rq1-results}

Table~\ref{tab:tool-results} reports functional (F), security (S), and all-pass (A) rates.
We discuss A, then F and S, and finally statistical tests.

\textbf{(1) All-pass.}
As in \S\ref{sec:empirical-setup}, all-pass is the main metric: generated RTL must pass both the functional and the security testbenches.
\tool{} has the highest average all-pass on every model.
Averaged over models, all-pass is $61.6\%$ for \tool{}, $49.6\%$ for SecV, and $51.4\%$ for RESCUE, i.e., gains of $+12.0$ and $+10.2$ points.
GLM-5.2 shows the largest gain ($63.0\%$ vs.\ $47.0\%$ SecV and $45.0\%$ RESCUE); DS-v4 shows the smallest ($57.8\%$ vs.\ $49.2\%$ and $53.3\%$), but the gain is still clear.
The same order holds by language.
Averaged over models, \tool{} reaches $66.8\%$, $67.6\%$, $51.0\%$, and $61.0\%$ all-pass on Verilog, SystemVerilog, VHDL, and Python; the better baseline reaches at most $56.5\%$, $57.8\%$, $43.9\%$, and $49.2\%$ on these languages.
VHDL is hardest for all methods, but \tool{} still improves all-pass in every model and language cell.

\textbf{(2) Functional pass rate.}
The largest gains are on F.
Averaged over models, F rises from $55.0\%$ (SecV) and $56.9\%$ (RESCUE) to $71.9\%$ (\tool{}), a $+15.0$ point gain over the stronger baseline.
\tool{} leads on every model average; GPT-5.6 reaches the highest F ($80.6\%$).
This matches the two-stage design in \S\ref{sec:rtl-gen}: Stage~1 builds a functional draft from the specification alone, and Stage~2 applies local obligation-guided edits.
SecV and RESCUE put retrieved security content into one generation pass, which more often lowers F.
This is the same trade-off seen under L1 and L2 in \S\ref{sec:empirical-results}.

\begin{table*}[t]
  \caption{Paired gains of \tool{} over SecV and RESCUE (percentage points).
  $\Delta$F/$\Delta$S/$\Delta$A: mean difference of five run-level rates.
  $p_A$: paired Wilcoxon signed-rank $p$-value on case-level mean all-pass ($n{=}392$);
  Overall uses run-level rates pooled across models ($n{=}25$).
  $^{\dagger}$ marks the non-significant security contrast discussed in the text.}
  \label{tab:sig-tests}
  \centering
  \begin{tabular}{@{}l *{3}{r} r *{3}{r} r@{}}
    \toprule
    & \multicolumn{4}{c}{vs.\ SecV} & \multicolumn{4}{c}{vs.\ RESCUE} \\
    \cmidrule(lr){2-5} \cmidrule(l){6-9}
    Model & $\Delta$F & $\Delta$S & $\Delta$A & $p_A$ & $\Delta$F & $\Delta$S & $\Delta$A & $p_A$ \\
    \midrule
    DeepSeek-v4 pro
      & $+11.6$ & $+5.8$ & $+8.6$ & $1.5{\times}10^{-6}$
      & $+9.4$ & $+1.4^{\dagger}$ & $+4.4$ & $8.5{\times}10^{-3}$ \\
    GLM-5.2
      & $+19.8$ & $+11.8$ & $+16.0$ & $5.2{\times}10^{-15}$
      & $+21.3$ & $+9.1$ & $+18.1$ & $5.0{\times}10^{-16}$ \\
    GPT-5.6 Luna
      & $+21.5$ & $+7.9$ & $+11.5$ & $1.6{\times}10^{-6}$
      & $+20.6$ & $+7.9$ & $+12.1$ & $1.4{\times}10^{-8}$ \\
    Mimo-v2.5 pro
      & $+17.0$ & $+12.2$ & $+13.9$ & $9.4{\times}10^{-14}$
      & $+10.9$ & $+5.4$ & $+7.2$ & $4.4{\times}10^{-6}$ \\
    MinMax-M3
      & $+14.5$ & $+6.8$ & $+10.0$ & $8.4{\times}10^{-7}$
      & $+12.8$ & $+4.4$ & $+8.9$ & $5.2{\times}10^{-6}$ \\
    \midrule
    Overall
      & $+16.9$ & $+8.9$ & $+12.0$ & $1.2{\times}10^{-5}$
      & $+15.0$ & $+5.6$ & $+10.2$ & $1.2{\times}10^{-5}$ \\
    \bottomrule
  \end{tabular}
\end{table*}

\textbf{(3) Security pass rate.}
\tool{} also has the highest average S on every model: $69.5\%$ versus $60.6\%$ (SecV) and $63.8\%$ (RESCUE), a $+5.6$ point gain over RESCUE.
The S gain is smaller than the F gain, which is expected because both baselines already add CWE knowledge.
Two cells break the pattern: on DS-v4 Verilog and SystemVerilog, RESCUE has higher S than \tool{} ($70.4\%$ vs.\ $68.8\%$; $81.0\%$ vs.\ $78.4\%$).
Even in these cells, all-pass still favors \tool{} ($55.1\%$ and $66.3\%$ vs.\ $48.2\%$ and $65.5\%$), because RESCUE loses more on F.
Higher S alone is therefore not enough for better joint success.

\textbf{(4) Statistical significance.}
Table~\ref{tab:sig-tests} lists paired gains of \tool{} over each baseline and the related tests.
For each model, $\Delta$F, $\Delta$S, and $\Delta$A are mean differences of the five run-level rates.
The $p_A$ column is a paired Wilcoxon signed-rank test on case-level mean all-pass over five runs ($n{=}392$).
The Overall row pools run-level rates across models ($n{=}25$) and uses a run-level Wilcoxon test.
All-pass gains are significant for every model against both baselines ($p_A \le 8.5{\times}10^{-3}$).
On the Overall row, all three metrics are significant ($p < 10^{-4}$).
Paired effect sizes on Overall all-pass are large (Cohen's $d{=}3.37$ vs.\ SecV; $d{=}1.91$ vs.\ RESCUE).

\begin{tcolorbox}[colback=gray!5, colframe=gray!60, boxrule=0.4pt, left=4pt, right=4pt, top=2pt, bottom=2pt]
\textbf{RQ1 Summary.}
\tool{} raises mean all-pass from $49.6\%$ (SecV) and $51.4\%$ (RESCUE) to $61.6\%$, with significant gains on F and S across models.
Retrieved or explored security knowledge helps, but does not replace signal-level obligations inferred from the specification.
\end{tcolorbox}

\subsection{RQ2: How Effective are the Components of \tool{}?}
\label{sec:rq2-components}

RQ1 shows that the full \tool{} pipeline outperforms SecV and RESCUE on all-pass.
We next isolate which design choices drive that gain.
We ablate three stages that realize goals G1--G3~(\S\ref{sec:overview}): symbolic obligation inference, LLM-as-Judge filtering, and two-stage generation.

\begin{figure*}[t]
  \centering
  \includegraphics[width=0.92\linewidth]{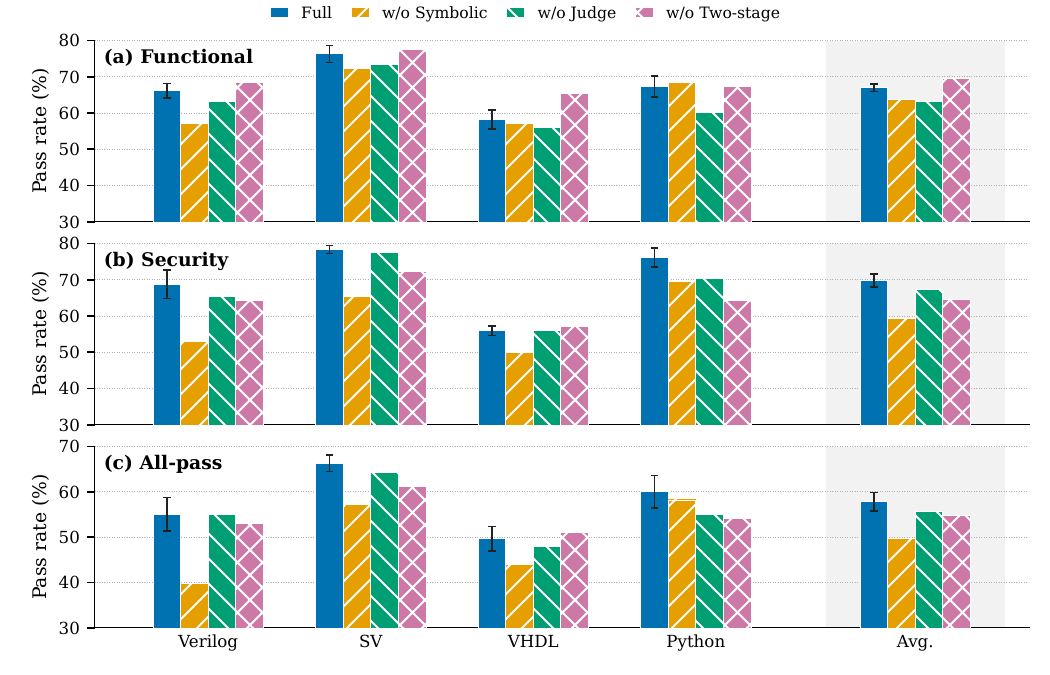}
  \caption{RQ2 component ablation on {\bench} (DeepSeek-v4-Pro).
  Panels show functional, security, and all-pass rates (\%) for Full \tool{} and three leave-one-out variants.
  Full reports the RQ1 five-run mean with error bars ($\pm$std); each ablation is one run.}
  \label{fig:rq2-ablation}
  \vspace{-0.5cm}
\end{figure*}

\subsubsection{Setup}
Because a full five-model ablation on {\bench} is costly, we run RQ2 with \textbf{DeepSeek-v4-Pro} as a representative backbone.
Among the five models in RQ1, it sits near the middle of the all-pass range, so it is a practical default for controlled component analysis.
We evaluate every ablation on the full benchmark (392 instances).
As in the agent study~(\S\ref{sec:disc-agents}), each ablation uses a single run; the Full row reuses the RQ1 five-run mean$\pm$std for the same model.
Metrics remain F, S, and A; Full is the reference in every comparison.

\subsubsection{Ablation Variants}
\label{sec:rq2-variants}

Each variant disables exactly one component and leaves the remaining pipeline unchanged:

\begin{itemize}[leftmargin=*,nosep]
  \item \textbf{w/o Symbolic} (\S\ref{sec:symbolic-infer}).
    We only remove the CWE-ontology matcher and ask the LLM to propose security obligations directly from the functional specification $s$, without a mitigation-evidence gap check. This variant tests whether deterministic symbolic inference (G1) is necessary, or whether free-form obligation proposal suffices inside the same downstream pipeline.
  \item \textbf{w/o Judge} (\S\ref{sec:obl-filter}).
    We skip applicability filtering and pass the full candidate set from symbolic matching to generation ($O'{=}O$).
    This variant tests the role of the precision gate that follows our high-recall matcher.
  \item \textbf{w/o Two-stage} (\S\ref{sec:rtl-gen}).
    We replace generate-then-revise with a single pass $c \sim G_\theta(s \,\|\, O')$, while retaining symbolic inference and the Judge.
    This variant tests whether separating a functional draft from obligation-guided revision (G3) is needed to avoid the functional penalty of security-augmented one-shot generation.
\end{itemize}

\subsubsection{Results}
\label{sec:rq2-results}

Figure~\ref{fig:rq2-ablation} reports F, S, and A for Full \tool{} and the three ablations.
We discuss all-pass first, then each removed component.

Full \tool{} reaches the highest average all-pass ($57.8\%$).
Removing symbolic matching hurts most (Avg.\ A $49.7\%$, $-8.1$ points).
Removing the Judge or two-stage generation yields smaller drops ($55.6\%$ and $54.8\%$).
The same order holds on Verilog, SystemVerilog, and Python; on VHDL, w/o Two-stage is slightly above Full on this single run ($51.0\%$ vs.\ $49.6\%$), but its average across languages remains below Full.
Overall, every component contributes positively to joint success, and symbolic inference is the largest driver.

\textbf{(1) w/o Symbolic.}
Without the ontology matcher, security falls from $69.8\%$ to $59.4\%$ Avg.\ S, and all-pass follows.
Functional rates stay close to Full ($63.8\%$ vs.\ $67.0\%$), so the damage is concentrated on security rather than on basic RTL synthesis.
This matches Finding~1--2 in \S\ref{sec:empirical}: when obligations are proposed by the LLM alone, defensive intent is incomplete even if the rest of the pipeline is unchanged.
Deterministic mitigation-evidence checking (G1) is therefore necessary for the security gains in RQ1.

\textbf{(2) w/o Judge.}
Skipping applicability filtering lowers Avg.\ A by $2.2$ points ($57.8\% \rightarrow 55.6\%$), with modest drops on both F ($67.0\% \rightarrow 63.3\%$) and S ($69.8\% \rightarrow 67.3\%$).
The Judge is thus a useful precision gate over the high-recall matcher, but not the main source of all-pass.
This is consistent with RQ3, where Judge over-filtering explains only a small share of Full-pipeline failures on DeepSeek-v4-Pro.

\textbf{(3) w/o Two-stage.}
One-shot generation $c \sim G_\theta(s \,\|\, O')$ lowers Avg.\ A to $54.8\%$ ($-3.0$ points), mainly through security ($69.8\% \rightarrow 64.5\%$).
Functional pass does not drop on this single run (Avg.\ F $69.6\%$ vs.\ Full $67.0\%$); the two-stage design still helps all-pass by letting obligation-guided revision specialize in security edits after a clean functional draft.
Together with the larger F gains of Full \tool{} over SecV/RESCUE in RQ1, this supports keeping generate-then-revise (G3), while the present single-run ablation indicates that its benefit on DeepSeek-v4-Pro appears more in joint success than in a large F penalty for one-shot prompting.

\begin{tcolorbox}[colback=gray!5, colframe=gray!60, boxrule=0.4pt, left=4pt, right=4pt, top=2pt, bottom=2pt]
\textbf{RQ2 Summary.}
Removing symbolic matching cuts Avg.\ all-pass from $57.8\%$ to $49.7\%$; removing the Judge or two-stage yields smaller drops ($55.6\%$ / $54.8\%$).
Symbolic inference is the primary contributor to security and all-pass; Judge filtering and two-stage generation provide complementary gains.
\end{tcolorbox}

\subsection{RQ3: Why Does \tool{} Still Fail?}
\label{sec:rq3-errors}

RQ1 shows that \tool{} beats SecV and RESCUE on all-pass, yet failures remain.
This RQ studies those failures: where the pipeline breaks, what tests fail, and what to improve next.

\subsubsection{Setup and classification}
We inspect all $9{,}800$ evaluations of \tool{} on {\bench} (5 models $\times$ 5 runs $\times$ 392 instances).
In total, $6{,}035$ pass all-pass and $3{,}765$ fail ($38.4\%$).
For each failure, we check the pipeline log and mark where it broke and how the tests failed.
We mark \textbf{Stage} as the first pipeline step that went wrong.
We mark \textbf{Outcome} as whether F, S, or both failed.
Table~\ref{tab:error-taxonomy} lists the labels.

\begin{table*}[t]
  \caption{Failure taxonomy for RQ3.
  Each failed evaluation gets one stage (earliest explainable pipeline step) and one outcome (test pattern).
  Counts are over all $3{,}765$ failures.}
  \label{tab:error-taxonomy}
  \centering
  \begin{tabular}{@{}llp{0.85\columnwidth}r@{}}
    \toprule
    Axis & Label & Definition & \#Fail \\
    \midrule
    \multirow{4}{*}{Stage}
      & FSG extraction
        & FSG parse fails; use secure fallback~(\S\ref{sec:gspec})
        & 8 \\
      & Symbolic matching
        & Gold CWE not triggered; no correct obligations~(\S\ref{sec:matching})
        & 799 \\
      & LLM-as-Judge
        & Candidates exist, but judge rejects all~(\S\ref{sec:obl-filter})
        & 203 \\
      & Obligation-guided gen.
        & Kept obligations reach Stage~2, yet RTL fails~(\S\ref{sec:rtl-gen})
        & 2{,}755 \\
    \midrule
    \multirow{3}{*}{Outcome}
      & Sec-only fail
        & $T_\mathit{func}$ pass; $T_\mathit{sec}$ fail
        & 1{,}010 \\
      & Func-only fail
        & $T_\mathit{func}$ fail; $T_\mathit{sec}$ pass
        & 771 \\
      & Both fail
        & Both $T_\mathit{func}$ and $T_\mathit{sec}$ fail
        & 1{,}984 \\
    \bottomrule
  \end{tabular}
\end{table*}

\subsubsection{Results}
\begin{figure*}[!t]
  \centering
  \includegraphics[width=0.92\linewidth]{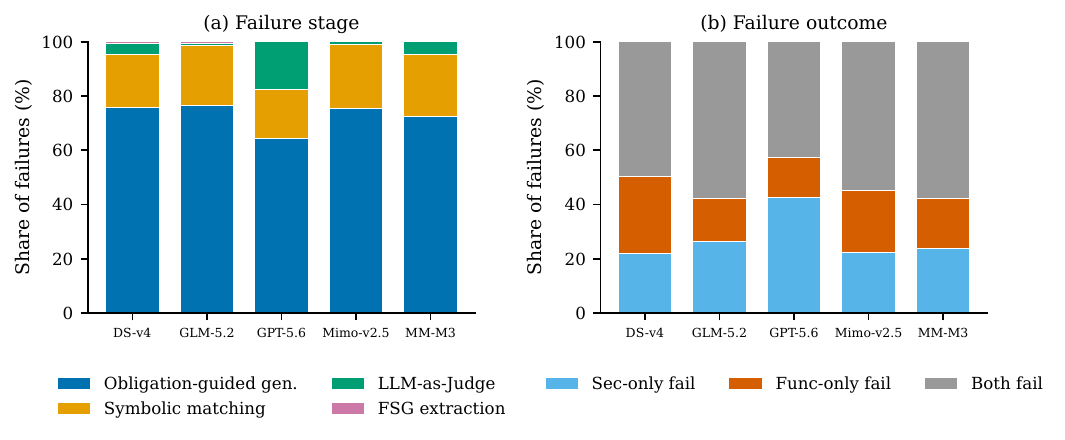}
  \caption{Error analysis of \tool{} failures on {\bench} ($n{=}3{,}765$).
  (a)~Earliest pipeline stage that explains the failure.
  (b)~Test outcome of the failure.}
  \label{fig:rq3-errors}
  \vspace{-0.5cm}
\end{figure*}

Figure~\ref{fig:rq3-errors} shows the share of each stage and outcome among failures, by model.

\textbf{(1) Failure stage.}
Of the $3{,}765$ failures, $2{,}755$ ($73.2\%$) keep obligations and still fail at obligation-guided generation.
Another $799$ ($21.2\%$) miss the gold CWE in symbolic matching, so Stage~2 never sees the right signal-level edits.
LLM-as-Judge over-filtering accounts for $203$ failures ($5.4\%$); FSG extraction accounts for only $8$ ($0.2\%$).
The pattern is stable across models: generation dominates for every backbone.
GPT-5.6 is the main outlier on the judge: $116$ of its $664$ failures ($17.5\%$) come from rejecting all candidates, versus under $5\%$ for the other models.

\textbf{(2) Failure outcome.}
Among the same $3{,}765$ failures, $1{,}984$ ($52.7\%$) fail both F and S, $1{,}010$ ($26.8\%$) fail security only, and $771$ ($20.5\%$) fail function only.
Both-fail is the largest class for four of five models.
GPT-5.6 again differs: $283$ security-only fails versus $282$ both-fails, and only $99$ functional-only fails, which matches its higher F in Table~\ref{tab:tool-results}.
So when GPT-5.6 fails, the RTL is more often functionally usable but still insecure.

\subsubsection{Implications}
These counts point to two concrete next steps.
First, Stage~2 revision must better enforce kept obligations without breaking function: this single stage covers $2{,}755$ failures.
Second, symbolic matching must raise gold-CWE recall: the $799$ matching misses never reach obligation-guided repair.
Judge filtering and FSG extraction are not the main bottlenecks today, except that stronger models may over-reject (GPT-5.6).
Overall, \tool{} already reduces the gap to prior secure-generation baselines; closing the rest mainly needs better obligation application and more complete CWE matching.

\begin{tcolorbox}[colback=gray!5, colframe=gray!60, boxrule=0.4pt, left=4pt, right=4pt, top=2pt, bottom=2pt]
\textbf{RQ3 Summary.}
Of $3{,}765$ \tool{} failures, $73.2\%$ keep obligations but fail at generation, and $21.2\%$ miss the gold CWE in matching.
LLM-as-Judge and FSG extraction cause few failures ($5.4\%$ and $0.2\%$).
Future work should focus on Stage~2 revision and CWE-matching recall.
\end{tcolorbox}

\section{Discussion}
\label{sec:discussion}

\subsection{Comparison with Coding Agents}
\label{sec:disc-agents}

We further compare \tool{} with coding-agent pipelines for RTL.
MAGE is a multi-agent system for functional RTL generation~\cite{zhao2025mage}: it samples candidates, generates testbenches, judges simulation feedback, and debugs iteratively.
We also evaluate Sec-MAGE, our security-aware variant that adds a Secure Testbench Agent to generate security-oriented tests and drive repair.

Because agent loops are expensive, we run this study only with DeepSeek-v4-Pro and a single run per method on the full {\bench} (392 instances).
For a fair cost comparison, we report \tool{} on the same model and the same first run.
Table~\ref{tab:agent-compare} summarizes pass rates and token cost.

\begin{table*}[t]
  \caption{Comparison with coding agents on {\bench} (DeepSeek-v4-Pro, one run).
  Per-language rates (\%) plus Avg.\ over the four languages.
  Token counts are totals over 392 instances.}
  \label{tab:agent-compare}
  \centering
  \begin{tabular}{@{}ll *{5}{c} r r r@{}}
    \toprule
    Method & & Verilog & SystemVerilog & VHDL & Python & Avg. & Calls & Tokens & Tokens/case \\
    \midrule
      & Functional & 72.5 & 81.6 & 74.5 & 59.2 & 71.9 & - & - & -\\
      & Security   & 15.3 & 16.3 & 12.2 & 12.2 & 14.0 & - & - & - \\
    \multirow{-3}{*}{MAGE}
      & All-pass   & 7.1  & 8.2  & 10.2 & 6.1  & 7.9
        & 3{,}116 & 13.7M & 34.9k \\
    \midrule
      & Functional & 76.5 & 84.7 & 72.5 & 58.2 & 73.0 & - & - &  -\\
      & Security   & 28.6 & 33.7 & 25.5 & 34.7 & 30.6 & - & - &  -\\
    \multirow{-3}{*}{Sec-MAGE}
      & All-pass   & 20.4 & 27.6 & 20.4 & 28.6 & 24.2
        & 5{,}178 & 33.2M & 84.6k \\
    \midrule
      & Functional & 65.3 & 79.6 & 57.1 & 69.4 & 67.9 & - & - &  -\\
      & Security   & 70.4 & 78.6 & 57.1 & 75.5 & 70.4 & - & - &  -\\
    \multirow{-3}{*}{\tool{}}
      & All-pass   & 55.1 & 69.4 & 51.0 & 59.2 & 58.7
        & 1{,}536 & 3.8M & 9.7k \\
    \bottomrule
  \end{tabular}
\end{table*}

On all-pass, \tool{} leads on every language (Table~\ref{tab:agent-compare}), with Avg.\ $58.7\%$ versus $7.9\%$ (MAGE) and $24.2\%$ (Sec-MAGE).
Security follows the same pattern (Avg.\ $70.4\%$ vs.\ $14.0\%$/$30.6\%$).
Functional correctness is mixed: MAGE and Sec-MAGE are higher on Verilog, SystemVerilog, and VHDL, while \tool{} is higher on Python; Avg.\ F is $71.9\%$/$73.0\%$/$67.9\%$.
Agent loops with a testbench agent help protect F, and Sec-MAGE further raises S and A over MAGE, but both remain far below \tool{} on all-pass at much higher cost.
A promising direction is to keep explicit security obligations for repair, and add a light testbench agent when F drops.

\textbf{Token cost.}
Token cost follows the same order as agent complexity (Table~\ref{tab:agent-compare}).
MAGE uses $13.7$M tokens ($34.9$k per case) and $3{,}116$ LLM calls.
Sec-MAGE uses $33.2$M tokens ($84.6$k per case) and $5{,}178$ calls.
\tool{} uses $3.8$M tokens ($9.7$k per case) and $1{,}536$ calls on the same model and run.
Relative to \tool{}, MAGE spends about $3.6\times$ more tokens and Sec-MAGE about $8.7\times$ more.
\tool{} therefore offers a better trade-off between security and functionality at substantially lower token cost than the adapted agent baselines under this cost-controlled setting.

\subsection{Threats to Validity}
\label{sec:threats}

\textbf{Internal validity.}
LLM sampling is stochastic.
For the main comparison we repeat each setting five times and report mean$\pm$std; the agent study uses one run for cost reasons, so those absolute rates are less stable.
We also re-implement SecV and adapt RESCUE and MAGE to {\bench}.
Faithful re-implementation reduces bias, but differences from the original artifacts may remain.

\textbf{External validity.}
{\bench} covers five resource-access CWE families chosen for MIHW priority and port-level observability.
These families are only part of hardware security: OpenTitan alone lists many countermeasures aimed at fault injection and side channels outside our threat model.
Results may not transfer to those properties, to closed-source SoCs, or to full-chip multi-IP integration.
We evaluate five frontier LLMs and four HDLs; newer models or other languages may shift absolute rates, though the obligation-inference design itself is model-agnostic.

\textbf{Construct validity.}
All-pass depends on our functional and security testbenches.
Black-box, port-observable checks avoid teaching to internal signals, but they may still miss some insecure behaviors.
We mitigate this with manual review of cases and golden implementations, and by checking that golden RTL largely passes the suites.

\section{Related Work}
\label{sec:related}

\subsection{LLM-Based RTL Generation}
A growing body of work applies LLMs to register-transfer-level (RTL) design.
Benchmarks such as VerilogEval~\cite{liu2023verilogeval}, RTLLM~\cite{lu2024rtllm}, and CVDP~\cite{pinckney2025comprehensive} evaluate functional correctness via simulation against golden designs.
Domain-adapted models and pipelines (including VeriGen~\cite{thakur2024verigen}, RTLCoder~\cite{liu2025rtlcoder}, ChipNeMo~\cite{liu2023chipnemo}, and multi-agent systems such as MAGE~\cite{zhao2025mage}) further raise pass rates through fine-tuning, synthetic data, or iterative generate--simulate--repair loops~\cite{yang2025large}.
Related tools also repair syntax errors in generated RTL~\cite{tsai2024rtlfixer}.
These lines of work treat functional behavior as the primary success criterion and do not systematically measure whether generated RTL satisfies security obligations.

\subsection{Security of LLM-Generated Code}
The security risks of LLM-generated software are well documented.
Pearce et al.~\cite{pearce2022asleep} find that roughly 40\% of Copilot suggestions in CWE-aligned scenarios are vulnerable, and datasets such as LLMSecEval~\cite{tony2023llmseceval} broaden evaluation over natural-language prompts.
Secure-generation methods then inject external knowledge: SafeCoder~\cite{he2024safecoder} security-tunes models for software, and RESCUE~\cite{shi2026rescue} retrieves CWE guidelines and sliced secure examples via RAG.
These studies assume a post-deployment patching model that silicon does not admit.

Hardware-facing work is more recent.
HardSecBench~\cite{chen2026hardsecbench} synthesizes Verilog and firmware-C tasks spanning many CWEs and reports that models often pass functional checks while failing security tests; concurrent audits observe similar residual-state and access-control issues~\cite{ibnat2025trusting}.
LLMs can also repair known hardware security bugs once the bug is identified~\cite{ahmad2024hardware}, but that setting starts from buggy RTL rather than from an obligation-omitted functional specification.
On the generation side, SecV~\cite{fan2025secv} retrieves Hardware-CWE knowledge-graph subgraphs, and SecFSM~\cite{hu2026secfsm} specializes this idea to FSM security for SoC control logic.
Outside the LLM setting, formal tools such as VeriSketch~\cite{gao2019verisketch} enforce security when properties are stated up front~\cite{athalye2022knox,mohr2023leave}, yet such intent is often incomplete or left implicit~\cite{dessouky2019hardfails,riaz2017implicit,rashid2016unknown}.

Compared to related work, {\bench} evaluates real SoC IP across four HDLs, and {\tool} symbolically infers signal-level obligations from a functional-semantic graph and a CWE ontology.

\section{Conclusion}
\label{sec:conclusion}

Insecure RTL cannot be patched after tape-out, yet LLM-based generators are still judged mainly by functional tests.
We release {\bench}, a multi-HDL benchmark of real SoC IP with black-box functional and security oracles under practice-motivated omission of security obligations.
Across five frontier models, vanilla generation is functionally strong but insecure; adding CWE knowledge helps, while unaided self-thinking cannot close the gap and security prompting trades away functionality.
\tool{} addresses this by symbolically inferring signal-level obligations from a functional-semantic graph and a CWE ontology, then applying them in a two-stage generation.

In this work, our scope is resource-access properties checked by black-box tests; side-channel and fault defenses typically need white-box oracles and fit repair-from-buggy-RTL settings better than pure generation from functional specs.
In the future, we plan to broaden the CWE ontology within this resource-access class, and to study repair workflows for weaknesses that demand white-box evaluation.

\section*{Acknowledgments}

This work was supported by National Key R\&D Program of China (No. 2024YFB4506400).
Guang Yang is also supported by the Postdoctoral Fellowship Program of CPSF under Grant Number GZC20260902.

\bibliographystyle{IEEEtran}
\bibliography{main}

\end{document}